\documentclass[aps,floatfix,twocolumn,letterpaper,showpacs]{revtex4}

\usepackage{amsmath}
\usepackage{amsfonts}
\usepackage{amssymb}
\usepackage{mathrsfs}
\usepackage{graphicx}
\usepackage{mathtools}

\usepackage{color}

\newcommand{\defeq}{\mathrel{\mathop:}=}

\begin{document}

\title{Nonequilibrium Dynamics of an Ultracold Dipolar Gas}
\author{A. G. Sykes}
\affiliation{JILA, University of Colorado and National Institute of Standards and Technology, Boulder, Colorado 80309-0440, USA}
\author{J. L. Bohn}
\affiliation{JILA, University of Colorado and National Institute of Standards and Technology, Boulder, Colorado 80309-0440, USA}

\begin{abstract}
We study the relaxation and damping dynamics of an ultracold, but not quantum degenerate, gas consisting of dipolar particles.  These simulations are performed using a direct simulation Monte Carlo method and employing the highly anisotropic differential cross section of dipoles in the Wigner threshold regime.  We find that both cross-dimensional relaxation and damping of breathing modes occur at rates that are strongly dependent on the orientation of the dipole moments relative to the trap axis.  The relaxation simulations are in excellent agreement with recent experimental results in erbium. The results direct our interest toward a less-explored regime in dipolar gases where interactions are dominated by collision processes rather than mean-field interactions. 

\end{abstract}

\maketitle

\section{Introduction}

Much of the attention on ultracold dipolar gases has heretofore focused on the quantum degenerate regime, where dipolar interactions can significantly influence the behavior of the gas through the mean-field.  Aspects of this influence include changing the shape and mechanical stability of the gas~\cite{CrNature2007,PfauCrStabilisationNature,CrStability1DLattice,RussBiss2,UweFischerPRA2006}, as well as altering the excitation spectrum to include low-energy roton modes in a Bose-Einstein condensate~\cite{Roton1,WilsonRotonPRL2008,CorsonBohnPRA,RussBissRotonPRL}.    A host of related phenomena have been predicted and observed~\cite{LahayeDipoleReview2009,TicknorAnisotropicSuperfluid,TicknorAnisotropicCoherence,KRbNature2013}, driven by the direct action of the long-ranged, anisotropic dipolar interaction on the particles' motion.

By contrast, gases at a slightly higher temperature behave more classically, and their mean-field energy is overcome by kinetic energy as the prime source of dynamics in the gas.  In such a situation the strength and anisotropy of the dipolar interactions can be made manifest through collisions, rather than through mean-field effects~\cite{JinBohnPRA2014}.  A very recent experiment showed this explicitly, finding that collisional relaxation of a gas of erbium atoms at $\sim 400$ nK occurred on time scales that varied by a factor of four, depending on the orientation of the atoms' magnetic dipole moments~\cite{FerlainoErXDR}.  This landmark result illustrates the potential for anisotropic dipolar scattering to profoundly influence the kinetics of a cold, thermal gas, from rethermalization and relaxation, to viscosity and the propagation of sound, to name but a few features.  

In this article we construct a model of the cold, nondegenerate dipolar gas by numerically solving the Boltzmann equation.  The model is based on the direct simulation Monte Carlo (DSMC) algorithm~\cite{BirdBook1,BirdBook2}, which is appropriate to the dilute limit found in experiments, when the mean-free path $\lambda_{\rm mf}$ of the atoms in the gas is comparable to, or larger than the characteristic scale $L$ of the gas (i.e., Knudsen number  $K\!\! n\equiv \lambda_{\rm mf}/L \gtrsim1$).  Using this model, we explore the thermal relaxation and damping of a dipolar gas that is suddenly taken out of equilibrium.  Where applicable, our results are in excellent agreement with the return to equilibrium of the erbium gas in Ref.~\cite{FerlainoErXDR}, and in particular describe the dependence of relaxation rate on polarization direction of the dipoles.  Further, we characterize the damping rate of breathing mode oscillations generated in the gas, finding that this damping is also strongly dependent on polarization, and is slower than the rethermalization rate. We also evaluate the relevance of mean-field interactions in the gas. Although the density of erbium in the cross-dimensional relaxation experiment was not sufficiently high to observe mean-field effects, we briefly discuss how to modify the DSMC method to include such physics (using {\it particle-in-cell} methods for a dipolar-Vlasov equation). 

The outline of this paper is as follows: In section~\ref{sec:Experiment} we introduce and discuss the details of a cross-dimensional rethermalization experiment which we will model. In section~\ref{sec:BoltzEqDSMC} we provide a very brief introduction to the Boltzmann equation and discuss its historical significance in statistical mechanics. Section~\ref{subsec:DSMC} outlines the basic features of our DSMC algorithm, and section~\ref{subsec:Differential} discusses the differential scattering cross sections for low-energy dipolar interactions. Section~\ref{sec:meanfield} discusses and quantifies the mean-field interaction in the gas. Section~\ref{sec:results} contains our results for fermions, and compares these results to experimental data. Section~\ref{sec:BosonResults} contains similar results, but for bosons. In section~\ref{sec:Conclusions} we draw conclusions and discuss possible avenues for future research.

\section{Cross-Dimensional Relaxation of a Dipolar Gas}\label{sec:Experiment}

For concreteness, we here contemplate the experimental situation of Ref.~\cite{FerlainoErXDR}.  We employ the notation of that experiment, and use the same values of trap frequencies, density, and species (erbium).  We stress, however, that the simulations can be made completely general for cold dipolar gas experiments in the thermal regime, including polar molecules.  

\begin{figure}
 \includegraphics[width=7cm]{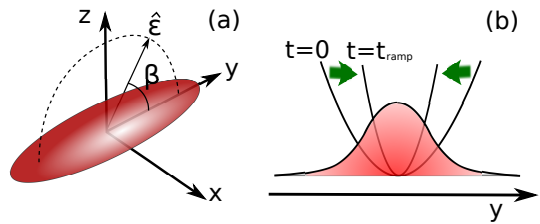}
\caption{(Color online) The initial state of the gas is shown in (a), where the atoms occupy the equilibrium state of an approximately cylindrically symmetric trap, elongated in the $y$ direction. The dipole alignment direction is given by $\hat{\boldsymbol\varepsilon}$. As depicted in (b), the experiment begins when the trap-frequency in the $y$ direction is {\it suddenly} increased, sending the system out of equilibrium. Rethermalisation dynamics depends on the angle $\beta$ between the dipole alignment direction and the $y$ axis.}\label{fig:ExpCartoon}
\end{figure}

Experiments involving cross-dimensional relaxation have a long history in cold atoms, going back to the work with caesium~\cite{MonroeCsXDR}. Other experiments include work on Bose-Fermi~\cite{GoldwinBoseFermiXDR} and Fermi-Fermi mixtures~\cite{FermiFermiXDR}. The experimental scenario  we consider is shown in Fig.~\ref{fig:ExpCartoon}. The gas begins in the equilibrium state of an approximately cylindrically symmetric trap, with the dipole alignment direction in the $y$--$z$ plane of the laboratory reference frame. The gas is weakly trapped in the $y$ direction, and tightly trapped in the $x$ and $z$ directions. The dipole alignment direction, $\hat{\boldsymbol\varepsilon}$, makes an angle $\beta$ with the $y$--axis.

Over a (fast) time scale $t_{\rm ramp}$, the trapping frequency along the $y$--axis is significantly increased, sending the system out of equilibrium. The atoms, whose distribution is initially still elongated along the $y$ direction, gain extra momentum along this direction (over the time-scale of a quarter trap-period). Rethermalisation requires the redistribution of this additional momentum/potential-energy in the $y$ direction into the $x$ and $z$ directions. Due to the highly anisotropic nature of the dipole-dipole interaction, the rate at which this rethermalization (redistribution) occurs depends strongly on the angle, $\beta$, between the dipole alignment direction and the $y$--axis (see Fig.~\ref{fig:ExpCartoon}).

This experiment was recently performed in Innsbruck~\cite{FerlainoErXDR}, as a very beautiful demonstration of the standards in precision and control over cold-atomic systems. The atomic species used was $^{167}$Er (a fermion), which has an exceptionally large magnetic dipole moment of $7\mu_{\rm B}$ where $\mu_{\rm B}$ is the Bohr magneton (compared to $^{87}$Rb with 1$\mu_{\rm B}$ and $^{52}$Cr with 6$\mu_{\rm B}$, $^{164}$Dy has 10$\mu_{\rm B}$). The experiment began with an initial temperature of 426nK. Relative to the density of the system, this corresponds to a regime;
 $\bar{n}\lambda_{\rm T}^3\approx0.25$,
where $\bar{n}$ is the average density in the trap, and $\lambda_{\rm T}=\hbar\sqrt{2\pi/(m k_{\rm B}T)}$ is the thermal de Broglie wavelength. In this sense, the system (although cold) is not {\it deeply} within a regime of quantum degeneracy. This then implies that the classical Boltzmann equation should provide the appropriate theoretical description. This being said, quantum-mechanical effects may indeed be a source of error in our simulations, and we attempt to quantify this in Section~\ref{subsec:QuantumManyBody}. 

In spite of this (relatively) low phase-space density, the system is still sufficiently cold such that the ratio between the thermal de Broglie wavelength, and a characteristic dipole-length scale; $a_{\rm d}=C_{\rm dd}m/(8\pi\hbar^2)$, (where $C_{\rm dd}=\mu_0\mu^2$ for magnetic dipoles, and $C_{\rm dd}=d^2/\epsilon_0$ for electric dipoles, $\mu_0$ and $\epsilon_0$ are respectively the permeability and permittivity of the vacuum, $\mu$ and $d$ are the magnetic and electric dipole moments) is;
 $\lambda_{\rm T}/a_{\rm d}\approx39$.
From this, we conclude that the two-body scattering physics is strongly within the quantum regime, and the differential scattering cross-sections are chosen accordingly~\cite{JinBohnPRA2014}.

\section{The Boltzmann Equation for dipolar gases and the DSMC Method}~\label{sec:BoltzEqDSMC}

\subsection{General considerations}

The ability to trap and cool atoms with large magnetic dipole moments, such as chromium~\cite{CrExpt1,CrExpt2}, dysprosium~\cite{DyExpt1,DyExpt2}, and erbium~\cite{ErExpt1} provide exciting possibilities for observing novel many-body states (for a recent example, see Ref.~\cite{RecentLevPaper}). Dipolar molecules are another source of potentially even stronger interactions in dipolar gases~\cite{KRbScience2008,KRbJapan,KRbNature2010,KRbNature2013,CsRbNagerl1,SrRbSchreck1}. Developing theoretical tools to understand dipolar gases is currently a very active area of research~\cite{Ticknor1,Ticknor2,BaillieBlakie1,CorsonBohnPRA,PRA-87-043620,BabadiDemlerTwoDimDipoles,NatuWilsonPRA2013}.

A particularly challenging task in many-body physics is to develop theoretical methods for treating out-of-equilibrium physics. To this end, we report on our progress towards a general tool for simulating out-of-equilibrium dynamics of the normal dipolar gas. Our approach is based on the Boltzmann equation, which we solve using the DSMC algorithm. The motivation for using DSMC typically occurs when the assumptions of fluid mechanics (which generally centres around some form of the Navier-Stokes equation) break down, and one must account for the granular nature of matter (usually, although not exclusively, via statistical mechanics). Bird's DSMC algorithm has evolved over recent decades into a remarkably versatile and useful tool which has been applied across seemingly disparate fields of research~\cite{DSMCapplications,CercignaniBoltzReview}. 

Stochastic particle methods, such as DSMC, have been applied to ultra-cold gases in a number of previous works. For instance, a variation of the method we describe here was used to study evaporative cooling enroute to Bose-condensation~\cite{Wu1EvapCooling}. In Ref.~\cite{Wade}, collisions between two thermal clouds near a $d$-wave resonance was simulated. The results compared very nicely to experiment~\cite{Otago_dwave_collisions}. Other examples include the study of collective modes in finite temperature dynamics~\cite{JacksonZaremba1,JacksonZaremba2,JacksonZaremba3}, sympathetic cooling of molecules~\cite{Barletta1}, and degenerate Fermi gas dynamics~\cite{TosiPRA2003}. To our knowledge, our work is the first time dipolar differential cross sections have been used~\cite{JinBohnPRA2014}. This reduces the efficiency of the DSMC by introducing a rejection-sampling algorithm to sample the differential cross sections. However, we find that numerical convergence is still easily attainable on standard commodity hardware. 

The classical Boltzmann equation describes the statistical mechanics of particles in a many-body system with two-body elastic collisions. Its modern derivation typically involves truncation of the BBGKY hierarchy~\cite{BoltzEqReview} such that two-body (and higher) distribution functions factorize into products of single-body distribution functions (this assumption was referred to by Boltzmann as the {\it stosszahlansatz}, or the assumption of molecular chaos). The equation for a single component gas reads
\begin{equation}\label{eq:BE1}
 \left[\frac{\partial}{\partial t}+\frac{{\bf p}}{m}.\nabla_{\bf r}+{\bf F}.\nabla_{\bf p}\right]f=C[f]
\end{equation}
where $f=f({\bf r},{\bf p};t)$ is the single particle phase-space distribution, i.e. $fd^3{\bf r}d^3{\bf p}$ is the expected number of atoms within the phase-space volume $({\bf r},{\bf p})\rightarrow({\bf r}+d^3{\bf r},{\bf p}+d^3{\bf p})$, $m$ is the particle mass, ${\bf F}$ denotes the external forces acting on the system, i.e ${\bf F}=-\nabla_{\bf r}U({\bf r},t)$ where $U({\bf r},t)$ is some external potential (trapping potential), and finally 
\begin{equation}\label{eq:CollisionIntegral}
 C[f]=\int \!\frac{d^3{\bf p}_1}{m}\int \!d\Omega \frac{d\sigma}{d\Omega}|{\bf p}-{\bf p}_1|\left[f^{\prime}f_1^{\prime}-ff_1\right]
\end{equation}
is the collision integral. We have used the common notation $f_1^{(\prime)}=f({\bf r},{\bf p}_1^{(\prime)};t)$. In principle, one may wish to include a mean-field contribution into the external potential. We discuss the relative importance of this mean-field term in section~\ref{sec:meanfield}, and demonstrate its insignificance for the purpose of simulating the experiment in Ref.~\cite{FerlainoErXDR}. 

The collision integral in Eq.~\eqref{eq:BE1} provides a mechanism for rethermalization via two-body collisions. Two particles (coinciding at the point ${\bf r}$) collide with momenta ${\bf p}$ and ${\bf p}_1$, and emerge from the collision with momenta ${\bf p}^\prime$ and ${\bf p}_1^\prime$. Net energy and momentum are conserved in the collision, meaning 
\begin{subequations}\label{eq:ConservationEquations}
  \begin{gather}
   {\bf P}={\bf P}^\prime\\
   |{\bf p}_{\rm rel}|=|{\bf p}_{\rm rel}^\prime|
  \end{gather}
\end{subequations}
where ${\bf P}^{(\prime)}=({\bf p}^{(\prime)}+{\bf p}_1^{(\prime)})/2$ and ${\bf p}^{(\prime)}_{\rm rel}={\bf p}^{(\prime)}-{\bf p}_1^{(\prime)}$ denote center-of-mass, and relative, momentum respectively. The differential cross-section 
\begin{equation}
 \frac{d\sigma}{d\Omega}=\frac{d\sigma}{d\Omega}({\bf p}_{\rm rel},{\bf p}_{\rm rel}^\prime)
\end{equation}
contains information regarding the likelihood of two particles colliding (given an incident relative momentum ${\bf p}_{\rm rel}$), and the likelihood of a post-collision relative momentum given by ${\bf p}_{\rm rel}^\prime$. Intriguingly, cross-sections which exhibit time-reversal symmetry, that is  $\frac{d\sigma}{d\Omega}({\bf p}_{\rm rel},{\bf p}_{\rm rel}^\prime)=\frac{d\sigma}{d\Omega}({\bf p}_{\rm rel}^\prime,{\bf p}_{\rm rel})$, yield irreversible dynamics in the Boltzmann equation as demonstrated by Boltzmann's famous $H$-theorem~\cite{BoltzmannsLegacy}. The relevant differential cross-section for dipolar particles has been derived and discussed in detail in a recent article~\cite{JinBohnPRA2014}, and we will briefly summarise the necessary results in section~\ref{subsec:Differential}. 

Analytic solutions to the Boltzmann equation are difficult to come by~\cite{BoltzmannRecentArxiv,BoltzmannsLegacy}. An important exception are the well known equilibrium solutions, the Maxwell-Boltzmann distribution;
\begin{equation}\label{eq:MB}
 f({\bf r},{\bf p};t)=f_{\rm MB}({\bf r},{\bf p})\defeq\frac{N}{Z}\exp\left[-\frac{p^2/2m+U({\bf r})}{k_{\rm B}T}\right],
\end{equation}
where $k_{\rm B}$ is Boltzmann's constant, $T$ is the temperature, $N$ is the total number of particles, and $Z=\int\!d^3{\bf r}d^3{\bf p}\exp\left[-\frac{p^2/2m+U({\bf r})}{k_{\rm B}T}\right]$ gives the correct normalisation. Using the conservation laws in Eq.~\eqref{eq:ConservationEquations}, it is straight-forward to see that $C[f_{\rm MB}]=0$, and $f_{\rm MB}$ is a stationary solution to Eq.~\eqref{eq:BE1}.

We wish to solve the Boltzmann equation~\eqref{eq:BE1} under the following dynamical scenario: Starting from an equilibrium initial distribution, Eq.~\eqref{eq:MB}, we change the trapping potential $U({\bf r},t)=\frac{1}{2}m\left[\omega_x^2 x^2+\omega_y(t)^2 y^2+\omega_z^2 z^2\right]$, where 
\begin{equation}\label{eq:trap}
 \omega_y(t)=\left\{
\begin{array}{ll}\vspace{0.1cm}
 \omega_y^{(0)}  & t<0\\ \vspace{0.1cm}
 \sqrt{1+s \frac{t}{t_{\rm ramp}}}\,\omega_y^{(0)}  & 0<t\leq t_{\rm ramp}\\ 
 \sqrt{1+s}\;\omega_y^{(0)} & t>t_{\rm ramp}
\end{array}
\right.
\end{equation}
such that, over the ramp-time $t_{\rm ramp}$, the trap frequency in the $y$-direction is changed by a factor $\sqrt{1+s}$. The choice of square-root dependence on time corresponds to linearly increasing the laser power in an optical dipole trap. The spatial anisotropy, created by the dipole-alignment direction, creates a bias for scattering into particular momentum states. This has implications for the rate of rethermalization, which becomes dependent on the angle between the $y$-axis and the dipole alignment direction. We investigate the rethermalization dynamics as a function of this angle, and compare it to the experimental work of Ref.~\cite{FerlainoErXDR}.

Attempting to solve the Boltzmann equation by discretizing the temporal axis and the phase-space dimensions is a futile exercise as (for all but the most trivial cases) one will always run out of computational resources, before numerical convergence is achieved. Viable alternatives to find an approximate solution in a close-to-equilibrium scenario do exist however. For instance, the so-called {\it method of moments} approach was used in Ref.~\cite{BabadiDemlerTwoDimDipoles} to study collective excitations of two-dimensional dipolar fermions in a perturbative limit. In Ref.~\cite{KavoulakisThermalRelaxationPRL} a variational method was employed to predict relaxation behaviour in $s$-wave interacting gases. Our method is more generally applicable to a wider variety of far-from-equilibrium scenarios, although it is more numerically intense than other methods.

\subsection{The DSMC method}\label{subsec:DSMC}

The starting point for DSMC approximates the distribution function $f$, by $N_{\rm T}$ test-particles each with position and momenta $[{\bf r}_i,{\bf p}_i]$ which are found by randomly sampling $f({\bf r},{\bf p};t=0)$. That is
\begin{equation}\label{eq:f1}
 f({\bf r},{\bf p};0)\approx\xi\sum_{i=1}^{N_{\rm T}}\delta({\bf r}-{\bf r}_i)\delta({\bf p}-{\bf p}_i)
\end{equation}
where $\xi=N/N_{\rm T}$ is the ratio of real-particles to test-particles. The goal is to force the test-particles to evolve in time $\left[{\bf r}_i(t),{\bf p}_i(t)\right]$ such that their relationship to $f$ shown in Eq.~\eqref{eq:f1}, remains true at all times. The computational complexity thus increases with $N_{\rm T}$.

On time scales, $\Delta t$, much shorter than the mean-collision-time, the evolution of each test particle is given by its classical trajectory in the potential. Assuming $\Delta t$ is also much shorter than the trap period, this is well approximated using a predictor-corrector (symplectic integrator) method,
\begin{subequations}\label{eq:CollisionlessEvolution}
\begin{align}
 {\bf q}_i&={\bf r}_i(t)+\frac{\Delta t}{2m}{\bf p}_i(t)\\
{\bf p}_i(t+\Delta t)&={\bf p}_i(t)+{\bf F}_i\Delta t\\
{\bf r}_i(t+\Delta t)&={\bf q}_i+\frac{\Delta t}{2m}{\bf p}_i(t+\Delta t),
\end{align}
\end{subequations}
where ${\bf F}_i=-\nabla_{{\bf q}_i}U({\bf q}_i,t)$ is the external force acting on the $i$-th test-particle. This is often referred to as the free-streaming dynamics. 
Note that, if the classical trajectory of a single particle in the trap can be solved analytically (which is obviously straight forward in the case of a harmonic potential), then Eqs.~\eqref{eq:CollisionlessEvolution} can be replaced by this analytic solution. This provides an advantage in that $\Delta t$ need not be small compared to the trap period (but still must remain small compared to the mean-collision-time). 
In effect, Eqs.~\eqref{eq:CollisionlessEvolution} account for the left hand side of the Boltzmann equation as shown in Eq.~\eqref{eq:BE1}. 

In order to include the effects of the collision integral [on the right hand side of Eq.~\eqref{eq:BE1}], a spatial grid is introduced, and the test-particles are binned into the volume-elements $\Delta V$ of this grid. This grid needs to be chosen carefully. The size of the volume-element effectively represents the finite resolution of the delta-function in our numerics. For this reason it needs to be small since all physical quantities will be coarse-grained over this volume-element. However, we will use the population of test-particles within each volume-element to stochastically check for collisions, and therefore, the volume-element must be large enough to contain multiple test-particles (in order to obtain reliable statistics). Being certain that one has the necessary combination of {\it large-enough} $N_{\rm T}$ and {\it small-enough} $\Delta V$ is an important numerical convergence test. 

Once the spatial grid has been established, we check $N_{\nu}(N_\nu-1)/2$ pairs of particles within the $\nu$th volume element ($N_{\nu}$ is the population of the $\nu$th volume element). In this step, the computational complexity acquires a $N_{\nu}^2$ dependence, and simulations will become unfeasible if individual volume elements contain too many test-particles. The collision probability is given by
\begin{equation}\label{eq:Pij}
 P_{ij}=\xi\frac{\Delta t}{m\Delta V} |{\bf p}_{\rm rel}|\sigma({\bf p}_{\rm rel})
\end{equation}
where ${\bf p}_{\rm rel}={\bf p}_{i}(t)-{\bf p}_{j}(t)$ and 
\begin{equation}
 \sigma({\bf p}_{\rm rel})=\int d\Omega_{{\bf p}_{\rm rel}^\prime}\frac{d\sigma}{d\Omega}({\bf p}_{\rm rel},{\bf p}_{\rm rel}^\prime)
\end{equation}
is the total cross section (as a function of relative momentum betwen particles $i$ and $j$), found by integrating the differential cross section over all solid angles of scattered relative momentum. Computational parameters must be chosen such that $P_{ij}\ll 1$. The collision proceeds if $R<P_{ij}$ where $R$ is a randomly generated number, with uniform distribution between 0 and 1. If the collision proceeds, we establish the post-collision relative momentum ${\bf p}_{\rm rel}^\prime$ by treating the differential cross-section $\frac{d\sigma}{d\Omega}$ as a probability distribution for ${\bf p}_{\rm rel}^{\prime}$, and stochastically sample it using a rejection-sampling algorithm (see Appendix~\ref{app:RejectionSampling} for more details). The center-of-mass momentum is conserved during the collision. In this way, at each time-step in our simulation, collisions are stochastically implemented, in correct accordance with the total-cross section, the local density, the local velocity distribution, and the differential scattering.

Our numerical algorithm described here, has some subtle inferiorities compared to certain other algorithms described in the literature. Deficiencies include the absence of locally-adaptive spatial grids (to efficiently account for dramatic variations in spatial density), scaled collision probabilities (without which the number of operations in the algorithm scales as $\sim\! N_{\rm T}^2$, rather than a potential $\sim\! N_{\rm T}$ scaling), and locally adaptive time steps~\cite{Wade,WadeThesis}. However, the cold atomic vapours under current consideration have relatively small numbers of particles, and we have thoroughly tested for, and found, excellent numerical convergence in all of our simulations. For this reason, we do not implement the complete set of modern sophistications within the DSMC.

\subsection{Differential scattering in dipolar gases}\label{subsec:Differential}

The cross-section formulae used in this work were derived in Ref.~\cite{JinBohnPRA2014} using the Born approximation for the scattering amplitude between two dipolar particles, with dipole moments aligned along an alignment direction $\hat{\boldsymbol\varepsilon}$ (we use $\hat{}$ to denote a unit vector). The formulae are $\frac{d\sigma_{\rm F,B}}{d\Omega}({\bf p}_{\rm rel},{\bf p}_{\rm rel}^{\prime})=a_d^2\left|g_{\rm F,B}({\bf p}_{\rm rel},{\bf p}_{\rm rel}^{\prime})\right|^2$, where
\begin{widetext}
 \begin{subequations}
 \begin{align}
g_{\rm F}({\bf p}_{\rm rel},{\bf p}_{\rm rel}^{\prime})&=\frac{1}{\sqrt{2}}\frac{4(\hat{{\bf p}}_{\rm rel}.\hat{\boldsymbol\varepsilon})(\hat{{\bf p}}_{\rm rel}^{\prime}.\hat{\boldsymbol\varepsilon})-2\left[
(\hat{{\bf p}}_{\rm rel}.\hat{\boldsymbol\varepsilon})^2+(\hat{{\bf p}}_{\rm rel}^{\prime}.\hat{\boldsymbol\varepsilon})^2
\right](\hat{{\bf p}}_{\rm rel}.\hat{{\bf p}}_{\rm rel}^{\prime})}{1-(\hat{{\bf p}}_{\rm rel}.\hat{{\bf p}}_{\rm rel}^{\prime})^2}\label{eq:gF}\\
g_{\rm B}({\bf p}_{\rm rel},{\bf p}_{\rm rel}^{\prime})&=\frac{1}{\sqrt{2}}\left[-2\frac{a}{a_d}-\frac{2(\hat{{\bf p}}_{\rm rel}.\hat{\boldsymbol\varepsilon})^2+2(\hat{{\bf p}}_{\rm rel}^{\prime}.\hat{\boldsymbol\varepsilon})^2-
4(\hat{{\bf p}}_{\rm rel}.\hat{\boldsymbol\varepsilon})(\hat{{\bf p}}_{\rm rel}^{\prime}.\hat{\boldsymbol\varepsilon})(\hat{{\bf p}}_{\rm rel}.\hat{{\bf p}}_{\rm rel}^{\prime})}{1-(\hat{{\bf p}}_{\rm rel}.\hat{{\bf p}}_{\rm rel}^{\prime})^2}+\frac{4}{3}\right],\label{eq:gB}
 \end{align}
\end{subequations}
\end{widetext}
$a_d$ is the dipole length scale given by $a_d=m\mu_0\mu^2/(8\pi\hbar^2)$, $\mu_0$ is the vacuum permeability, $\mu$ is the atoms magnetic dipole moment ($\mu=7\mu_{\rm B}$ in the case of erbium), and $a$ is the s-wave scattering length. 
The subscripts ${\rm F}$ and ${\rm B}$ respectively correspond to fermionic and bosonic symmetry constraints ($^{167}$Er which was used in the experiment~\cite{FerlainoErXDR} is fermionic).

The total cross section, which we use to evaluate the collision probability in Eq.~\eqref{eq:Pij} can also be evaluated analytically~\cite{JinBohnPRA2014},
\begin{subequations}\label{eq:TotalCrossSect}
\begin{align}
 \sigma_{\rm F}({\bf p}_{\rm rel})&=a_d^2\frac{\pi}{3}\left[3+18\cos^2(\eta)-13
\cos^4(\eta)\right]\\
\sigma_{\rm B}({\bf p}_{\rm rel})&=a_d^2\frac{\pi}{9}\Big\{72a^2-24a\left[1-3\cos^2(\eta)\right]\nonumber\\
 &\quad\quad+11-30\cos^2(\eta)+27\cos^4(\eta)\Big\}
\end{align}
\end{subequations}
where $\eta=\cos^{-1}(\hat{{\bf p}}_{\rm rel}.\hat{\boldsymbol\varepsilon})$ is the angle between the dipole alignment direction and the incoming relative momentum. Equation~\eqref{eq:TotalCrossSect} [(a) or (b) depending on whether the collision pair are identical fermions or bosons] is used in Eq.~\eqref{eq:Pij} to evaluate the likelihood of a collision.

Once it has been established whether or not the collision occurs, the post-collision relative velocity is found by sampling the distribution function
\begin{equation}\label{eq:ProbDistFun}
 P_{\rm F,B}(\theta_{\rm rel},\phi_{\rm rel};\eta)=\frac{1}{\sigma_{\rm F,B}({\bf p}_{\rm rel})}\frac{d\sigma_{\rm F,B}}{d\Omega}({\bf p}_{\rm rel},{\bf p}_{\rm rel}^{\prime})\sin\theta_{\rm rel}.
\end{equation}
Note that we only need to sample $\theta_{\rm rel}$ and $\phi_{\rm rel}$ since $\eta$ is given to us by the (already known) incoming relative momentum of the collision pair. 
The {\it collision-reference-frame} $(x_{\rm cf},y_{\rm cf},z_{\rm cf})$ is defined such that the $z_{\rm cf}$-axis points along the direction of ${\bf p}_{\rm rel}$, and the dipole-alignment direction $\hat{\boldsymbol\varepsilon}$ lies in the $x_{\rm cf}$--$z_{\rm cf}$ plane. The purpose of defining, and operating within the collision-reference-frame is to make the analytic formulae of Eqs.~\eqref{eq:gF} and~\eqref{eq:gB} as wieldy as possible. The coordinates $\theta$ and $\phi$ in Eq.~\eqref{eq:ProbDistFun} are the polar and azimuthal angles (respectively) of ${\bf p}_{\rm rel}^{\prime}$ in the collision-reference-frame. We (arbitrarily) decide to include the factor $\sin\theta$ into the definition of the probability distribution function (rather than the metric) such that $\int_0^{2\pi}\!d\phi\int_0^{\pi}\!d\theta \,P_{\rm F,B}(\theta,\phi;\eta)=1$. Sampling the probability distribution in Eq.~\eqref{eq:ProbDistFun} is not simple, so we use a rejection sampling algorithm which we describe in Appendix~\ref{app:RejectionSampling}.

To convert between the lab-reference-frame and the collision-reference-frame, we find
\begin{align}
 {\bf e}_1^{\rm cf}=&\left[\cos(\gamma)\cos(\phi_{\rm rel})\cos(\theta_{\rm rel})-\sin(\gamma)\sin(\phi_{\rm rel})\right]{\bf e}_1^{\rm lf}\nonumber\\
&\quad+\left[\cos(\gamma)\sin(\phi_{\rm rel})\cos(\theta_{\rm rel})-\sin(\gamma)\cos(\phi_{\rm rel})\right]{\bf e}_2^{\rm lf}\nonumber\\
&\quad-\cos(\gamma)\sin(\theta_{\rm rel}){\bf e}_3^{\rm lf}\\
 {\bf e}_2^{\rm cf}=&\left[-\sin(\gamma)\cos(\phi_{\rm rel})\cos(\theta_{\rm rel})-\cos(\gamma)\sin(\phi_{\rm rel})\right]{\bf e}_1^{\rm lf}\nonumber\\
&\quad+\left[\cos(\gamma)\cos(\phi_{\rm rel})-\sin(\gamma)\sin(\phi_{\rm rel})\cos(\theta_{\rm rel})\right]{\bf e}_2^{\rm lf}\nonumber\\
&\quad+\sin(\gamma)\sin(\theta_{\rm rel}){\bf e}_3^{\rm lf}\\
{\bf e}_3^{\rm cf}=&\cos(\phi_{\rm rel})\sin(\theta_{\rm rel}){\bf e}_1^{\rm lf}+
\sin(\phi_{\rm rel})\sin(\theta_{\rm rel}){\bf e}_2^{\rm lf}+\nonumber\\
&\quad\cos(\theta_{\rm rel}){\bf e}_3^{\rm lf}
\end{align}
where the angle
\begin{align}
  \gamma=&{\rm acot}\big\{\!\cos(\theta_{\rm rel})\cot(\phi_{\rm \varepsilon}-\phi_{\rm rel})\nonumber\\
&\quad-{\rm cot}(\theta_{\rm \varepsilon}){\rm csc}(\phi_{\rm \varepsilon}-\phi_{\rm rel})\sin(\theta_{\rm rel})\big\}
\end{align}
and 
\begin{align}
 \hat{{\bf p}}_{\rm rel}=&\sin(\theta_{\rm rel})\cos(\phi_{\rm rel}){\bf e}_1^{\rm lf}+\sin(\theta_{\rm rel})\sin(\phi_{\rm rel}){\bf e}_2^{\rm lf}\nonumber\\
&\quad+\cos(\theta_{\rm rel}){\bf e}_3^{\rm lf},\\
\hat{{\boldsymbol\varepsilon}}=&\sin(\theta_{\rm \varepsilon})\cos(\phi_{\rm \varepsilon}){\bf e}_1^{\rm lf}+\sin(\theta_{\rm \varepsilon})\sin(\phi_{\rm \varepsilon}){\bf e}_2^{\rm lf}+\cos(\theta_{\rm \varepsilon}){\bf e}_3^{\rm lf}.\label{eq:epsilonAngles}
\end{align}
We have used the common notation where ${\bf e}_{1,2,3}^{\rm lf,cf}$ denote the standard (unit) basis vectors of Euclidean space in either the lab- (lf) or collision- (cf) frame. The symbols $\theta_{\rm \varepsilon}$ and $\phi_{\rm \varepsilon}$ refer to the azimuthal and polar angles (respectively) of the dipole alignment direction in the lab frame, as shown in Eq.~\eqref{eq:epsilonAngles}.

\section{Discussion on the mean-field interaction}\label{sec:meanfield}

In a more general situation the inclusion of a mean-field interaction may be desirable~\cite{GueryOdelinMeanField1,GueryOdelinMeanField2}. This requires an alteration to the Boltzmann equation~\eqref{eq:BE1} such that ${\bf F}=-\nabla U({\bf r},t)$ now consists of two parts, $U({\bf r},t)=U_{\rm ext}({\bf r},t)+U_{\rm mf}({\bf r},t)$, an external potential $U_{\rm ext}$ and a mean-field potential $U_{\rm mf}$. Such an approach may be dubbed a dipolar-Vlasov equation in recognition of its similarity to the Vlasov equation used in plasma physics~\cite{Vlasov}. The mean-field potential is a dynamical variable (away from equilibrium) found from the convolution 
\begin{equation}\label{eq:Umf}
 U_{\rm mf}({\bf r},t)=\int d^3{\bf r}^\prime n({\bf r}^\prime,t)V_{\rm dd}({\bf r}-{\bf r}^\prime)
\end{equation}
where $n({\bf r},t)=\int\! d^3{\bf p}\;f({\bf r},{\bf p};t)$ is the spatial number density and $V_{\rm dd}({\bf r})$ is the dipolar interaction between two particles separated by ${\bf r}$. This is given by
\begin{equation}
 V_{\rm dd}({\bf r})=\frac{C_{\rm dd}}{4\pi}\frac{1-3(\hat{\bf r}\cdot\hat{\boldsymbol{\varepsilon}})^2}{r^3}.
\end{equation}
In general it is certainly true that the physics associated with the mean-field interaction can have a strong influence.

Upon including the mean field potential, the effects of interactions manifest within two distinct terms of the Boltzmann equation. The natural question arises whether or not there is some error akin to double counting due to the presence of both these terms. The collision term describes an instantaneous collision between exactly two particles within the gas, such that momenta is exchanged between these two particles. This effect is entirely local, and occurs irrespective of the other particles in the gas. On the other hand, the mean field consists of a collective effect due to every single particle in the gas. In this sense the two terms are conceptually distinct from one another. Serious problems begin to occur when the mean field interaction energy becomes particularly significant (taking up a large fraction of the total energy in the gas). In such a situation, the collisions can begin to occur, not on the background of a translationally invariant potential energy landscape (as it is generably assumed~\cite{JinBohnPRA2014}) but rather on an appreciably varying potential energy landscape, caused by the mean field of nearby particles. These problems arise when typical values of $na_d^3$  approach or exceed unity.  As we show below, this is not the case in our current realm of interest.

In order to ascertain the relevance of the mean field in Eq.~\eqref{eq:Umf} for our current simulation, we wish to consider the total mean-field energy per particle $e_{\rm mf}$ in the gas, and compare this to the temperature. That is, we calculate 
\begin{equation}
e_{\rm mf}=\frac{1}{2N}\int \! d^3 {\rm r}\;n({\bf r},t)U_{\rm mf}({\bf r},t).
\end{equation}
We are only interested in placing an approximate upper-bound on the value of $e_{\rm mf}$, so we simplify the situation at hand by assuming the density of the gas (at any given time) is given by a gaussian distribution with cylindrical symmetry about the dipole-alignment direction which (solely for the purpose of this discussion) we assume to be along the $z$-axis;
\begin{equation}
 n({\bf r})=\frac{N}{(2\pi)^{3/2}\sigma_{\perp}^2\sigma_{z}}\exp\left[-\frac{x^2+y^2}{2\sigma_{\perp}^2}-\frac{z^2}{2\sigma_z^2}\right].
\end{equation}
One could perform a more realistic calculation in the absence of cylindrical symmetry, but analytic calculations are difficult in this case. Although a numerical solution is not difficult, it only changes the result by a factor of order unity, and is therefore not of interest to us at this stage. The wonderfully elegant Fourier transform of $V_{\rm dd}({\bf r})$ allows for the analytic calculation of $e_{\rm mf}$~\cite{LahayeDipoleReview2009}
\begin{equation}
 e_{\rm mf}=-\frac{N}{48\sqrt{\pi^3}}\frac{C_{\rm dd}}{\sigma_{\perp}^2\sigma_z}h\left(\frac{\sigma_{\perp}}{\sigma_z}\right)
\end{equation}
where 
\begin{equation}
 h(x)=\frac{1+2x^2}{1-x^2}-\frac{3x^2\textrm{arctanh}\sqrt{1-x^2}}{(1-x^2)^{3/2}}
\end{equation}
is a function generally of order unity (although $h(1)=0$ since the angular average of $V_{\rm dd}$ is zero). In an attempt to draw some broad conclusions, we simply consider the prefactor in $e_{\rm mf}$ and compare it to the temperature:
\begin{equation}
 \kappa=\frac{1}{k_{\rm B}T}\frac{N}{48\sqrt{\pi^3}}\frac{C_{\rm dd}}{\sigma_{\perp}^2\sigma_z}.
\end{equation}
In the experiment of Ref.~\cite{FerlainoErXDR} which we are currently interested in, the quantity $\kappa$ is never more than $\kappa\lesssim0.02$, indicating that physics associated with the mean-field is likely to be insignificant, at least to a first level of approximation. 

In other situations (involving higher densities, or larger dipole length scales), where $\kappa$ becomes appreciably large, incorporating the mean-field into the simulation may be necessary. The computational issues of doing so are, to a certain extent, manageable (see for example the vast literature on {\it particle-in-cell} methods  used to solve the ordinary Vlasov equation in the field of plasma physics~\cite{ParticleInCell}). Briefly, the process involves binning the particles in position space to find the density $n({\bf r},t)$, smoothing the density via convolution with a suitably chosen gaussian kernel, and then calculating the potential, using Eq.~\eqref{eq:Umf}, and ultimately the force ${\bf F}$~\cite{ParticleInCell}. For issues relating to clarity, we currently wish to relegate further details of this procedure to a future publication.

\section{Results for Fermions}\label{sec:results}

The choice of physical parameters in our simulation are taken directly from Ref.~\cite{FerlainoErXDR}. These are
\begin{align}
&N=8\times10^4\quad &\textrm{total atom number}\nonumber\\
&T=426{\rm nK}\quad &\textrm{initial temperature}\nonumber\\
&m=2.77\times10^{-25}{\rm kg}\quad &^{167}\textrm{Er mass}\nonumber\\
&a_d=5.25{\rm nm}\quad &^{167}\textrm{Er dipole length scale}\nonumber\\
 &
\begin{array}{ll}
  \omega_x=2\pi\times393{\rm Hz}\\
  \omega_y^{(0)}=2\pi\times38{\rm Hz}\\
  \omega_z=2\pi\times418{\rm Hz}\\
 \end{array} & \Bigg\}\textrm{initial trap}\nonumber\\
&t_{\rm ramp}=14{\rm ms}\quad &\textrm{ramp time}\nonumber\\
&s=1.8\quad &\textrm{final trap, $y$--axis [see Eq.~\eqref{eq:trap}].}\nonumber
\end{align}
We vary the computational parameters $N_{\rm T}$ and $\Delta V$ until numerical convergence is achieved. This has typically occurred when $N_{\rm T}\approx N$, although we perform our simulations right through to $N_{\rm T}= 4\times N$ to thoroughly check the convergence. We find these simulations converge rather rapidly with $\Delta V$~\cite{BillingsleyProbabilityTheory}, however we perform simulations right through to $\bar{n}\Delta V=0.35$ (where $\bar{n}$ is the initial trap-averaged density), with $N_{\rm T}= 4\times N$, to be certain of convergence.

\subsection{Anisotropic pseudo-temperature}

\begin{figure*}
 \includegraphics[width=14cm]{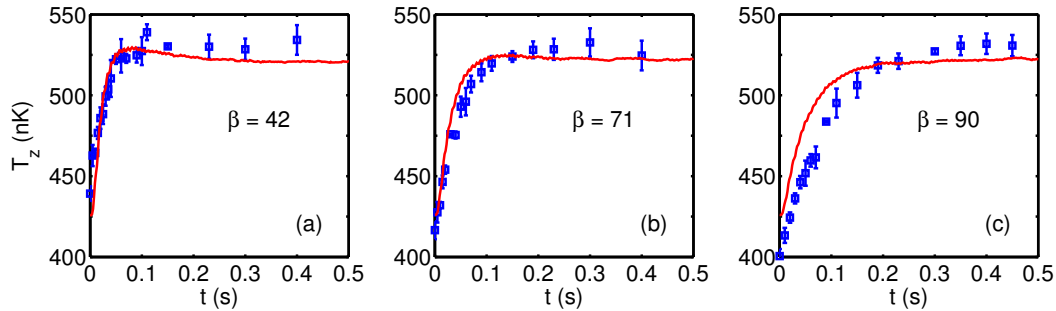}
\caption{(Color online) A comparison between the experimentally measured rethermalization process versus the results from the DSMC simulation. In each figure, the red solid line shows the result of the DSMC simulation, calculated analagously to Eq.~\eqref{eq:Tx}, but along the $z$ axis. The experimental data points are shown in blue with error bars. The agreement is reasonable, especially considering there was no post-processing made of the experimental data, nor any adjustments to the theory in order to produce these fits (no free parameters).}
\label{fig:ExpComp1}
\end{figure*}

\begin{figure*}
 \includegraphics[width=16cm]{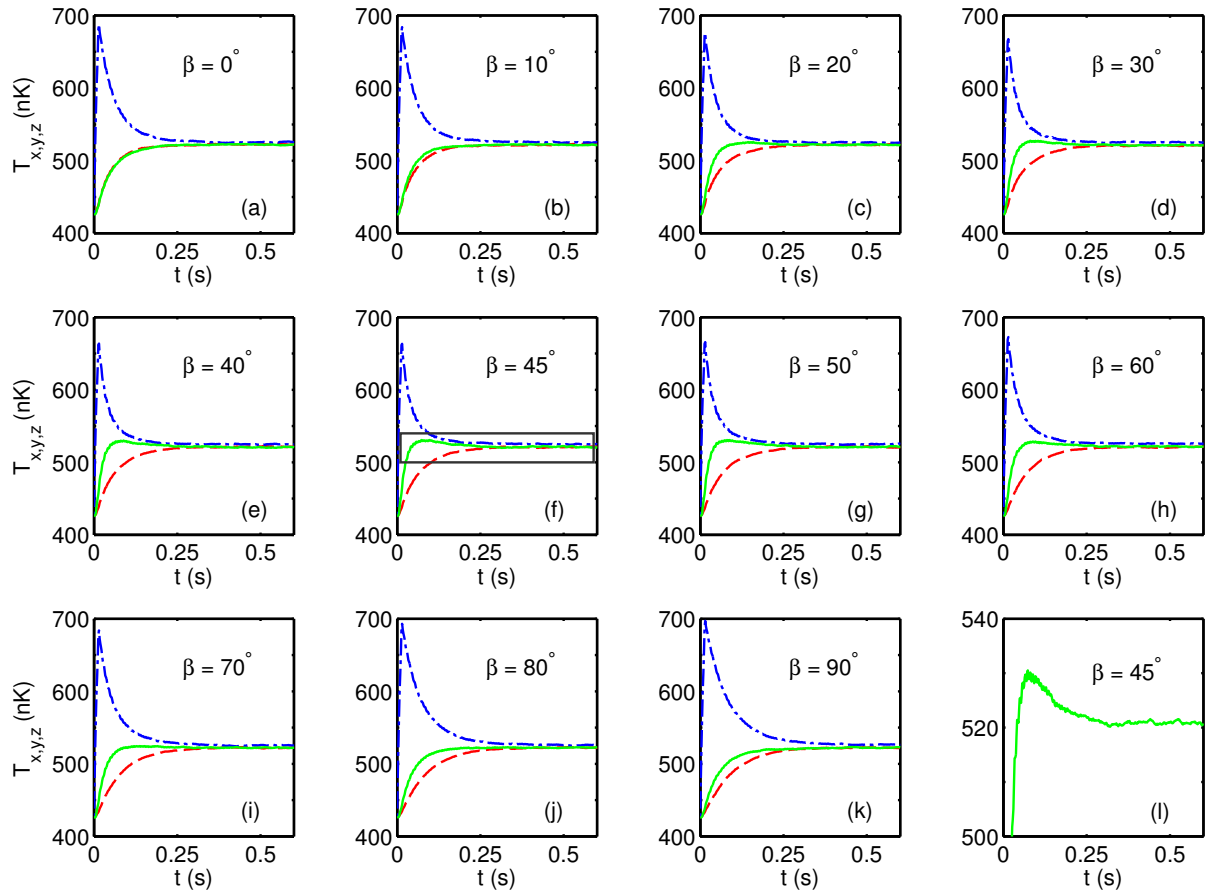}
\caption{(Color online) Shows the pseudo-temperatures along the $x$, $y$, and $z$ axes (shown in red-dashed, blue-dot-dashed, and green-solid lines respectively) defined analogously to Eq.~\eqref{eq:Tx}. In the first 14ms (the ramp time), the temperature along the $y$--axis increases as a result of the rapid change in the trap-frequency along this direction. After 14ms, the system relaxes down towards the new equilibrium state. The rate at which  $T_{\rm x}$, $T_{\rm y}$ and $T_{\rm z}$ return to an equilibrium value displays a strong dependence on $\beta$, which we explore further in section~\ref{subsec:RateOfRetherm}. An unexpected feature we observe is the non-monotonic path by which $T_{\rm z}$ returns to equilibrium near $\beta=45^{\circ}$. The effect is shown in greater detail in (l) by zooming in on the relevant part of (f). This is certainly an interesting consequence of the anisotropic dipole differential scattering, but note that the behaviour only occurs along one of the coordinate axes (the $z$-axis in this case) and, overall there is {\it no} violation of Boltzmann's $H$-theorem.}
\label{fig:TxTyTzVsTime}
\end{figure*}

To evaluate the rate of rethermalization, we find the standard-deviations of the test-particle distributions; for instance
\begin{align}
 \sigma_{x}(t)=\sqrt{\frac{1}{N_{\rm T}}\sum_{i=1}^{N_{\rm T}}x_i(t)^2},\quad
\sigma_{p_x}(t)=\sqrt{\frac{1}{N_{\rm T}}\sum_{i=1}^{N_{\rm T}}p_{xi}(t)^2},\nonumber
\end{align}
and equally for the $y$ and $z$ directions. We note that, a gaussian distribution provides a reasonably accurate approximation to the instantaneous empirical distribution of test particles in the simulation. However, the moments above are well defined, regardless of whether this is the case or not. From these standard deviations, we can define a time-dependent, anisotropic {\it pseudo-temperature}, related to the widths of the test-particle distribution function in each direction, relative to the instantaneous value of the trapping parameters, for instance;
\begin{align}
 \mathcal{T}_x=\frac{m\omega_x^2\sigma_x^2}{k_{\rm B}}{\rm , }\qquad
 \mathcal{T}_{p_x}=\frac{\sigma_{p_x}^2}{mk_{\rm B}},
\end{align}
and equally for the $y$ and $z$ axes. This definition makes particular sense in the case of a gaussian distribution. The two quantities; $\mathcal{T}_x$ and $\mathcal{T}_{p_x}$ above, can be combined into a single pseudo-temperature in the $x$--direction (or in any direction) given by the mean;
\begin{equation}\label{eq:Tx}
 T_{x}=\frac{\mathcal{T}_x+\mathcal{T}_{p_x}}{2}.
\end{equation}
The results of this analysis for the temperature along the $z$--axis is shown in Fig.~\ref{fig:ExpComp1}, along with the experimental data of Ref.~\cite{FerlainoErXDR}. 
A more complete set of results, for the temperatures in all three directions is shown in Fig~\ref{fig:TxTyTzVsTime}. An interesting observation we make is the apparent non-monotonic rethermalization behaviour of $T_{\rm z}$ near $\beta=45^{\circ}$ (this behaviour seems to exist right through $30^{\circ}\lesssim\beta\lesssim60^{\circ}$). This behaviour was not observed in the experiment, likely due to the fact that it is a subtle effect which may be difficult to measure. Indeed we note in Fig.~\ref{fig:ExpComp1} (a), the scatter and error bars in the experimental data points appear to be of a similar size, or even larger than the magnitude of the non-monotonic {\it hump} in the theoretical result.

\subsection{Analyzing the rate-of-rethermalization as a function of $\beta$}\label{subsec:RateOfRetherm}

\begin{figure}
 \includegraphics[width=8cm]{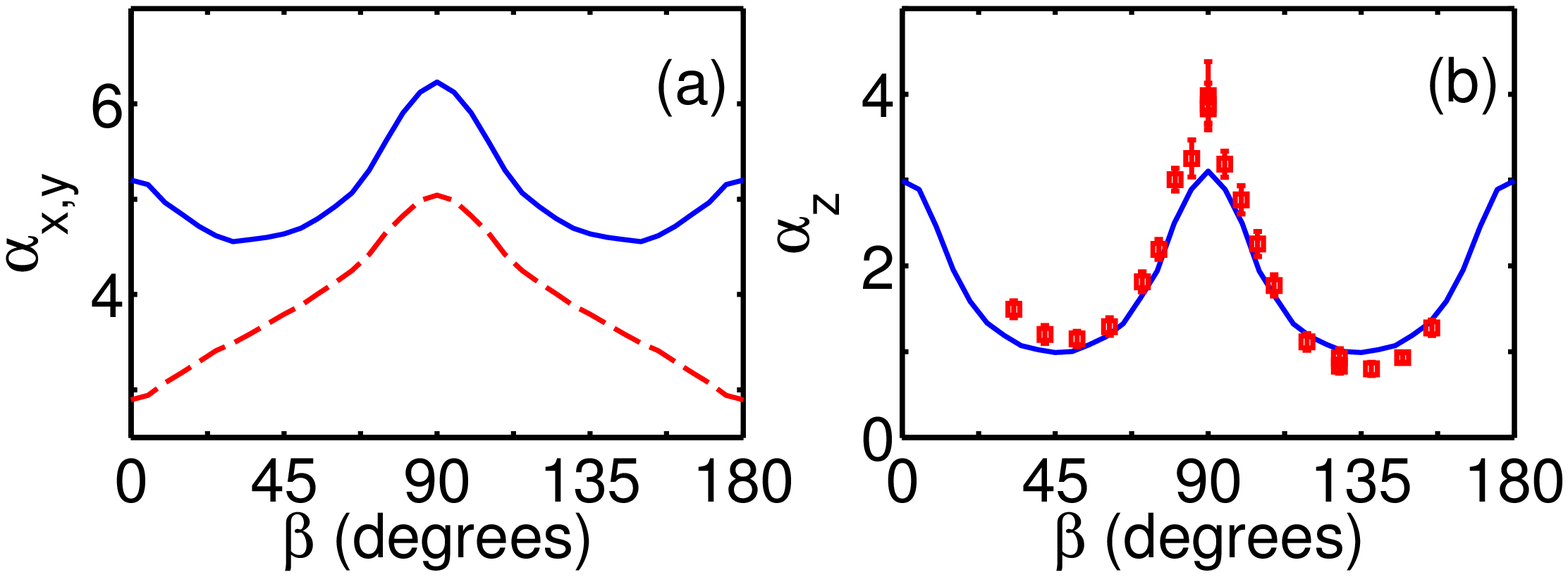}
\caption{(Color online) Shows $\alpha$ (the number of collisions required for rethermalization) as a function of the angle $\beta$ along the; (a) $x$ (red dashed line) and $y$ (blue solid line) directions, and (b) $z$ direction. The data shown in (b) is taken from the experiment in Ref.~\cite{FerlainoErXDR} (data was not taken in the $x$ and $y$ directions).}
\label{fig:alpha}
\end{figure}

In order to define a rate-of-rethermalization it is customary to fit an exponential decay curve to the equilibration dynamics shown in Figs.~\ref{fig:ExpComp1} and \ref{fig:TxTyTzVsTime}. For example, in the $z$-direction, one would write $T_{\rm z}(t)=T_{\rm z}^{\rm (eq)}+\Delta T_{\rm z}e^{-t/\tau_z}$, where $T_{\rm z}^{\rm (eq)}$ (a fit parameter) is the equilibrated temperature, and $T_{\rm z}^{\rm (eq)}+\Delta T_{\rm z}$ is the initial temperature (426nK in our case). The time-constant of this exponential decay curve, $\tau_z$, is then written as
\begin{equation}
 \tau_z=\frac{\alpha_z}{\bar{n}\bar{\sigma}_{\rm F,B}\bar{v}},
\end{equation}
where $\bar{v}=\sqrt{16k_{\rm B}T/\pi m}$ is the mean-velocity in the gas, and $\bar{\sigma}_{\rm F,B}$ is the total cross-section of Eq.~\eqref{eq:TotalCrossSect} averaged over all solid angles of the incoming relative momentum ${\bf p}_{\rm rel}$, such that $\bar{\sigma}_{\rm F}=(32\pi/15)a_d^2$ and $\bar{\sigma}_{\rm B}=8\pi a^2+(32\pi/45)a_d^2$.  In this way, the quantity $\bar{n}\bar{\sigma}_{\rm F,B}\bar{v}$ represents the mean-collision-frequency in the gas, and the quantity $\alpha$ can be conceptually thought of as the {\it number of collisions required for rethermalization}. The exact same procedure can be applied to the $x$ and $y$ axes. In our current situation $\alpha_{x,y,z}$ will be a function the angle $\beta$ between the dipole-alignment direction and the $y$ axis. The results, which agree well with experimental data from Ref.~\cite{FerlainoErXDR}, are shown in Fig.~\ref{fig:alpha}.

It should be noted that Refs.~\cite{FerlainoErXDR,JinBohnPRA2014} compute $\alpha_z$ in a simpler way, by approximating the short-time behaviour of the dynamics via the Enskog equation~\cite{Reif}. This has also shown adequate agreement with the data, but gives considerably less detail than the present DSMC simulations.

\subsection{Trap-oscillations and covariances in position and momentum space}

\begin{figure}
\includegraphics[width=8cm]{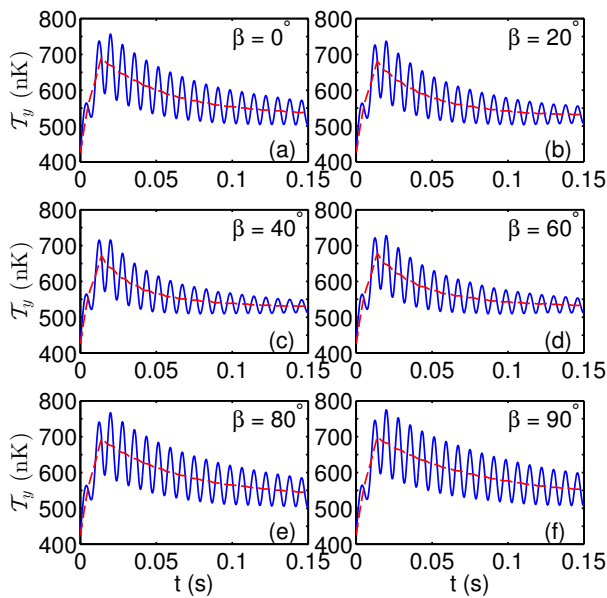}
 \caption{(Color online) The rapid change in trapping frequency along the $y$-axis generates a large breathing mode along this direction. This is shown above in plots of $\mathcal{T}_{y}$ versus time for a variety of different values of $\beta$. These breathing modes exist also in the momentum distribution, $\mathcal{T}_{p_y}$, and look identical to the plots above except that the oscillations are exactly $\pi$-radians out of phase (leading to the monotonic behaviour in $T_y$ shown in Fig.~\ref{fig:TxTyTzVsTime}). The dashed (red) line in each of the figures is $T_y$. We use Eq.~\eqref{eq:TrapOscFit} as a fit to the decay of this breathing mode. The breathing mode dynamics along the $x$ and $z$ axes are barely noticeable in our simulations.} \label{fig:TrapOscillations}
\end{figure}

\begin{figure}
\includegraphics[width=8cm]{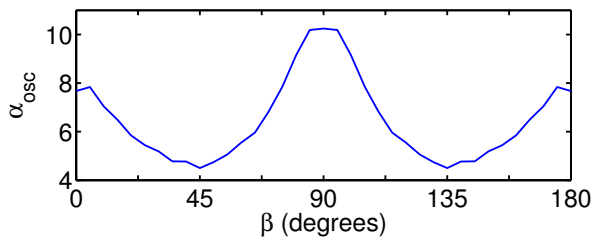}
 \caption{The breathing mode along the $y$-axis is damped over a time-scale $\tau_{\rm osc}$ found from Eqs.~\eqref{eq:TrapOscFit} and \eqref{eq:tauosc}. The dependence on $\beta$ is shown above. Note the qualitative similarity $\alpha_{\rm osc}$ (shown above) has to $\alpha_y$ in the blue line of Fig.~\ref{fig:alpha} (a). However, the oscillations take considerably longer to damp than the envelope, as $\alpha_{\rm osc}>\alpha_y$.}\label{fig:alpha_osc}
\end{figure}

The sudden change in the trap frequency along the $y$-axis gives rise to a breathing-mode along this direction (see Ref.~\cite{GueryOdelin1999} for a discussion of this subject in the case of a classical gas with hard-sphere interactions). The oscillations are apparent in either the position variable $\mathcal{T}_{y}$, or the momentum variable $\mathcal{T}_{p_y}$, but {\it not} in the sum $T_y$ which is plotted in Fig.~\ref{fig:TxTyTzVsTime} (since $\mathcal{T}_{p_y}$ and $\mathcal{T}_{y}$ oscillate exactly out of phase with each other). This behaviour is shown in Fig.~\ref{fig:TrapOscillations}. The experiment of Ref.~\cite{FerlainoErXDR} neither reported, nor searched for, any evidence of these oscillations or their damping periods (data was only analyzed along the $z$-axis). The frequency of the breathing mode is $2\omega_y(t>t_{\rm ramp})$~\cite{CercignaniBoltzReview,BoltzmannRecentArxiv}. Collisions will eventually cause this mode to damp out (intriguingly though, monopole modes are undamped in spherically symmetric harmonic traps). In order to quantify this, we subtract off the pseudo-temperature (shown by the red-dashed line in Fig.~\ref{fig:TrapOscillations}), and fit a decaying sinusoid to the data;
\begin{equation}\label{eq:TrapOscFit}
 \mathcal{T}_y(t)-T_y(t)\approx A e^{-t/\tau_{\rm osc}}\sin\left[\omega t+\delta\right].
\end{equation}
In the current experimental scenario the erbium gas lies firmly within the collisionless limit (trap frequency is significantly higher than the mean-collision frequency), and therefore the oscillation frequency is $\omega=2\sqrt{1+s}\;\omega_y^{(0)}$ i.e. twice the final trap frequency. Of course, if instead the experiment were in the hydrodynamic regime, rather than the collisionless regime, this would not be the case~\cite{GueryOdelin1999,JacksonPRL2002}. We only fit to the region $t>t_{\rm ramp}$ when the trap is no longer changing. The parameters $A$, $\tau_{\rm osc}$, and $\delta$ are all fitting parameters. We then scale the time-constant $\tau_{\rm osc}$ by the collision-frequency to give us 
\begin{equation}\label{eq:tauosc}
 \tau_{\rm osc}=\frac{\alpha_{\rm osc}}{\bar{n}\bar{\sigma}_{\rm F,B}\bar{v}}
\end{equation}
such that we can loosely interpret $\alpha_{\rm osc}$ as the {\it number of collisions required for the breathing mode to damp out}. Naively one might expect this to be the same as the $\alpha$ in section~\ref{subsec:RateOfRetherm}, and indeed we find distinct similarities, however the breathing mode takes considerably longer to damp out (a factor of 2 or more). The results for how $\alpha_{\rm osc}$ depends on $\beta$ is shown in Fig.~\ref{fig:alpha_osc}, note the qualitative similarity between Fig.~\ref{fig:alpha_osc} and the blue line in Fig.~\ref{fig:alpha} (a). We do not find that the other fitting parameters $A$ and $\delta$ have any significant dependence on $\beta$. However, $A$ does depend on the size of the perturbation to the trap, and $\delta$ depends on the ramp time $t_{\rm ramp}$ (this is apparent in the instantaneous quench, for which analytic formulae are straight-forward).

In contrast, breathing modes along the $x$ and $z$ axis are considerably less pronounced~\cite{YouMurray}. This is simply due to the fact that the perturbing force on the system in this situation is entirely along the $y$ axis (see Fig.~\ref{fig:ExpCartoon}).

If the quench were performed instantaneously, a simple analytic solution is available in the extreme-collisionless limit:
\begin{align}
 f({\bf r},{\bf p},t)=&f_{\rm MB}^{\rm (2D)}\left[(x,z),(p_x,p_z)\right]\times\nonumber\\
&\mathcal{M}\exp\left[-\frac{1}{2}(y\quad p_y){\boldsymbol{\Phi}}(t)^{-1}
\left(
\begin{array}{l} y \\ p_y
\end{array}
\right)\right]
\end{align}
where $f_{\rm MB}^{\rm (2D)}$ is the 2D Maxwell-Boltzmann distribution (along the $x$ and $z$ axes), $\mathcal{M}$ is a normalisation constant, and the covariance matrix 
\begin{equation}
 \boldsymbol{\Phi}=\left(
\begin{array}{cc}
 \zeta & \eta \\
 \eta & \theta
\end{array}
\right)
\end{equation}
is such that $\zeta=\langle y^2\rangle-\langle y\rangle^2$, $\eta=\langle y p_y\rangle-\langle y\rangle\langle p_y\rangle$, and $\theta=\langle p_y^2\rangle-\langle p_y\rangle^2$. Note that $\zeta$ and $\theta$ are proportional to the pseudo-temperatures $\mathcal{T}_y$ and $\mathcal{T}_{p_y}$ respectively, where as $\eta$ is the covariance between position and momentum space. Ignoring collisions in the system, these variances evolve according to~\cite{YouMurray};
\begin{subequations}\label{eq:Murray}
\begin{align}
 \zeta&=\frac{\zeta_0}{2}\left[1+\Gamma+(1-\Gamma)\cos\left(2\omega_y^{\rm (f)}t\right)\right]\\
\eta&=\frac{\sqrt{\zeta_0\theta_0}}{2}\left[\Gamma^{1/2}-\Gamma^{-1/2}\right]\sin\left(2\omega_y^{\rm (f)}t\right)\\
\theta&=\frac{\theta_0}{2}\left[1+\Gamma^{-1}+(1-\Gamma^{-1})\cos\left(2\omega_y^{\rm (f)}t\right)\right]
\end{align}
\end{subequations}
where $\zeta_0=k_{\rm B}T/[m(\omega_y^{(0)})^2]$, and $\theta_0=k_{\rm B}Tm$ are the initial spatial and momentum variances (respectively), and $\Gamma=\left(\omega_y^{(0)}/\omega_y^{\rm (f)}\right)^2$ is the ratio of initial-to-final trap frequencies (squared).

\begin{figure}
 \includegraphics[width=8cm]{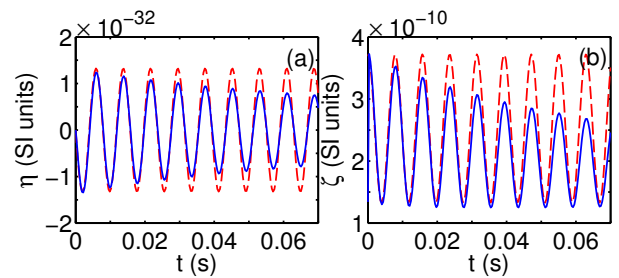}
\caption{(Color online) 
Comparison between the DSMC simulation for an instantaneous quench and the analytic formulae in Eqs.~\eqref{eq:Murray}. In (a) we plot the covariance between position and momentum space, and in (b) we plot the variance in position space (proportional to $\mathcal{T}_y$). This particular data is for a dipole alignment direction of $\beta=0$. The DSMC simulation is shown by the solid (blue) line, the analytic formulae by the dashed (red) line. The analytic formulae do an excellent job of correctly predicting the amplitude and phase of the oscillations. For this ratio of collision-to-trap frequency, the damping becomes appreciable on the order of several trap periods.
}\label{fig:CompareToMurray}
\end{figure}

We have performed simulations of the cross-dimensional relaxation procedure in the case of an instantaneous quench. The results are shown in Fig.~\ref{fig:CompareToMurray}, where we compare the simulation data to the analytic formulae of Eqs.~\eqref{eq:Murray}. The simulations reveal the increasing importance of collision-induced damping for times beyond several trap periods. The decay rate of the covariance $\eta$ depends on the dipole angle $\beta$. To within the numerical accuracy of these simulations, we find that the rate at which $\eta$ decays, and the dependence this decay has on $\beta$, is extremely close to that for $\zeta$ and $\theta$ (the pseudo-temperatures) shown in Fig.~\ref{fig:alpha_osc}.

\subsection{Quantum many-body effects}\label{subsec:QuantumManyBody}
\begin{figure}[ht!]
 \includegraphics[width=8cm]{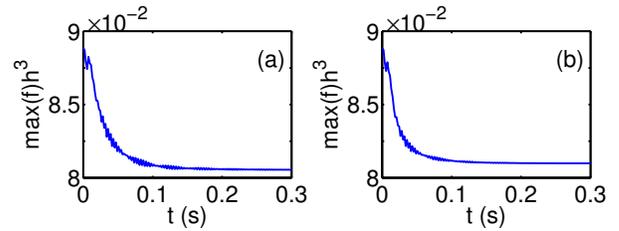}
\caption{Shows the (maximum) number of particles in a volume element of phase space equal to $h^3$ as a function of time for two separate dipole-alignment angles; (a) $\beta=0$, and (b) $\beta=45$. The phase space density decreases as the system equilibrates to a higher final temperature. From this, we estimate that quantum many-body effects are indeed small enough to be neglected (at least as a first approximation).}
\label{fig:PhaseSpaceDensityPlot}
\end{figure}

The Boltzmann equation, as written in Eq.~\eqref{eq:BE1}, treats the many-body dynamics of the system entirely in terms of classical mechanics. For our comparison with the experiment in Ref.~\cite{FerlainoErXDR}, this may conceivably be a source of error. In 1928, Nordheim made adjustments to the Boltzmann equation to account for the quantum-mechanical effects of Fermi-blocking and Bose-enhancement~\cite{Nordheim}. The net result of Nordheim's work was an alteration to the collision integral:
\begin{align}
 &C_{\rm N}[f]=\int \!\frac{d^3{\bf p}_1}{m}\int \!d\Omega \frac{d\sigma}{d\Omega}|{\bf p}-{\bf p}_1|\times \nonumber\\
&\left[f^{\prime}f_1^{\prime}\left(1\pm h^3\! f\right)\left(1\pm h^3\! f_1\right)
 \! -\! ff_1\left(1\pm h^3\! f^\prime\right)\left(1\pm h^3\! f_1^\prime\right)\right]
\label{eq:Nordheim}
\end{align}
where $h$ is Planck's constant, and the $+$ sign applies to identical bosons (Bose enhancement) while the $-$ sign applies to identical fermions (Fermi blocking). From this point of view, the quantum many-body effects in the system are determined by the phase-space density (see Ref.~\cite{GoulkoFermionincBoltzmann} for a discussion, and recent results, on the fermionic gas). Specifically how many particles occupy a volume of phase space equal to $h^3$. If this number is much less than one, quantum effects should be small, if this number is comparable to one, quantum effects will be important. The maximum phase-space density is plotted in Fig.~\ref{fig:PhaseSpaceDensityPlot} as a function of time for two different values of $\beta$. From this, we conclude that quantum many-body effects will have a negligible effect on the dynamics at this temperature. This goes some way in explaining the reasonably good agreement between our theory and experiment in this case. We do not expect our theory to provide quantitative accuracy at significantly lower temperatures, although modifying our algorithm to account for the mechanism of Bose-enhancement/Fermi-blocking is a future goal of this project. Speculating further on this, we note that the Boltzmann-Nordheim equation will have, not only a (potentially) different path to equilibrium, but also (at lower temperatures) a different equilibrium state as well (the famous Bose-Einstein and Fermi-Dirac distributions). How this would affect the dependence of $\alpha_{x,y,z}$ on $\beta$ is an interesting and open question.

\section{Results for Bosons}\label{sec:BosonResults}

It is very straightforward to repeat these simulations for a system of bosons simply by replacing $g_{\rm F}$ with $g_{\rm B}$ in the differential scattering cross-section and $\sigma_{\rm F}\rightarrow\sigma_{\rm B}$ (see Eqs.~\eqref{eq:gF}, ~\eqref{eq:gB}, and \eqref{eq:TotalCrossSect} in section~\ref{subsec:Differential}). We choose to keep the geometry of the trap, the atomic species, and the number of particles the same as that which was used in section~\ref{sec:results} for fermions. We set the $s$-wave scattering length $a=0$, to emphasize the peculiarities of the anisotropic dipolar differential scattering. The distinctions between bosonic versus fermionic scattering behaviour naturally alters details of the rejection sampling algorithm (see appendix~\ref{app:RejectionSampling}) and changes the results, but there is no conceptual change in what we are doing, so we provide less detail than we did for fermions. In addition, experimental data does not yet exist for bosons, so we cannot make the same comparisons in that respect.

\begin{figure}
 \includegraphics[width=8cm]{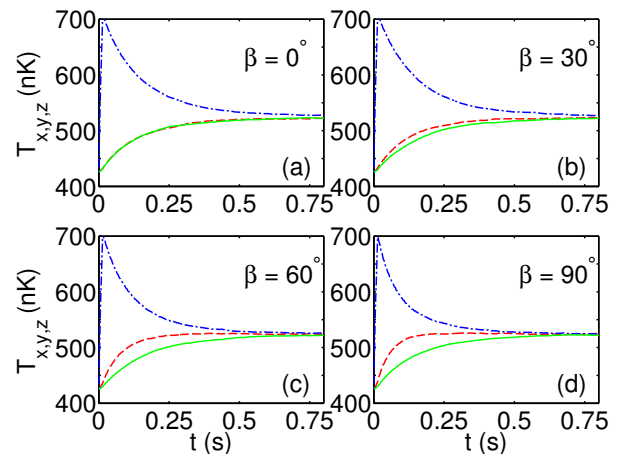}
\caption{(Color online) Shows the pseudo-temperatures along the $x$, $y$, and $z$ axes (shown by red-dashed, blue-dot-dashed, and green-solid lines respectively) as a function of time for the bosonic dipole scattering cross-section. Experimental data has not been taken for this case, but we observe the rethermalization rates showing a strong dependence on $\beta$ (particularly along the $x$-axis).}
\label{fig:BosonXDR}
\end{figure}

\begin{figure}
 \includegraphics[width=8cm]{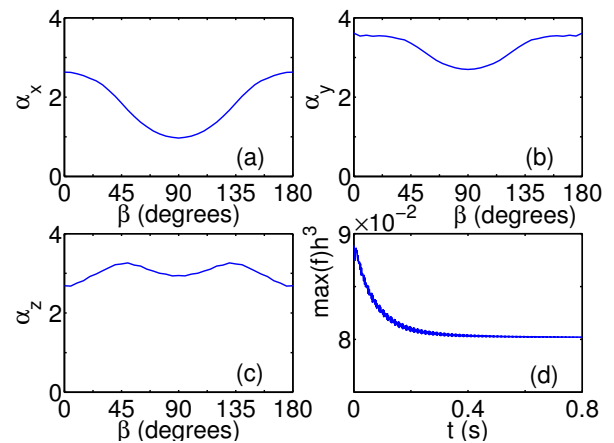}
\caption{(a), (b), and (c) show $\alpha_{x}$, $\alpha_y$, and $\alpha_z$ (the number of collisions required for rethermalization) as function of $\beta$ in the case of bosons. (d) shows the maximum phase space density as a function of time for $\beta=30^{\circ}$.}
\label{fig:BosonAlpha}
\end{figure}

\begin{figure}
 \includegraphics[width=8cm]{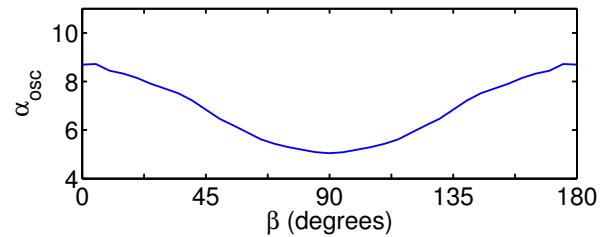}
\caption{The decay of the breathing mode along the $y$-axis in the case of bosons. Again, there is a strong qualitative similarity between this curve and the curve in Fig.~\ref{fig:BosonAlpha} (b), but an important quantitative difference in that $\alpha_{\rm osc}$ is larger by approximately a factor of two.}
\label{fig:AlphaOscBoson}
\end{figure}

Figure~\ref{fig:BosonXDR} shows the rethermalization of the pseudo-temperatures for bosons (analagous to Fig.~\ref{fig:TxTyTzVsTime} for fermions). Somewhat ironically, in the context of low-energy scattering, the rethermalization procedure takes approximately three times longer for bosons than for fermions with the same density and dipole-moment. This is due to the factor of three difference (for $a=0$) between the angularly averaged total cross sections $\bar{\sigma}_{\rm F}$ and $\bar{\sigma}_{\rm B}$~\cite{JinBohnPRA2014}. Increasing the $s$-wave scattering length $a$ would naturally change this situation. The nature of the differential cross-sections are such that a nonmonotonic rethermalization process is not observed for bosons [as it was in Fig.~\ref{fig:TxTyTzVsTime} (l)]. Figure~\ref{fig:BosonAlpha} (a)--(c) shows the {\it number of collisions required for rethermalization} as a function of $\beta$. In (d) we show the maximum phase-space density as a function of time for the case $\beta=30^{\circ}$. Again this indicates that the Boltzmann equation should provide an approximately accurate theoretical description at these densities and temperatures. Figure~\ref{fig:AlphaOscBoson} shows the {\it number of collisions required to damp out the breathing mode}. Note the qualitative similarity between $\alpha_{\rm osc}$ in Fig.~\ref{fig:AlphaOscBoson}, and $\alpha_{y}$ in Fig.~\ref{fig:BosonAlpha} (b), but with a quantitative difference of approximately a factor of two.

\section{Conclusions and Discussion}\label{sec:Conclusions}

In this article we have developed a DSMC numerical algorithm to solve the Boltzmann equation for an ultra-cold dipolar gas. We have used this method to study the cross-dimensional relaxation dynamics of a dipolar gas via a full simulation of the phase-space dynamics. Where applicable, we have compared our numerical results with the experimental data of Ref.~\cite{FerlainoErXDR} and found favourable agreement. This suggests that the DSMC algorithm provides a quantitative method for understanding the normal component in a dipolar gas. This is a promising result. The method is suitable for both fermions and bosons, although experimental data currently exists only for fermions. The method and results direct our interest toward a new regime where interactions in the gas manifest from collisions rather than the mean-field. 

More specifically, we have studied the damping of trap breathing modes in the system and quantified the pronounced dependence of rethermalization on the dipole-alignment direction. We find the breathing mode takes significantly longer (approximately a factor of two) to decay than the envelope for rethermalization, which is found by averaging over momentum-space and real-space dynamics. 

Our current work is entirely focused on the thermal gas, above quantum degeneracy. There are several reasons why understanding this normal component of an ultra-cold dipolar gas is important. For instance, attractive interactions along the dipole alignment direction (due to the mean-field) can destabilise the system~\cite{PfauCrStabilisationNature,SantosPRL2000,UweFischerPRA2006,CorsonPRA2013}. Thermal energy can counter-act this instability~\cite{RussBiss2,TicknorDipFinTemp}, therefore we expect the normal component to have a qualitative, as well as quantitative, role in the dynamics. Our method presented here, if combined/coupled with one of the many low-temperature theories (e.g.~\cite{cfieldreview,TomBillamFiniteTemp}) would constitute a complete finite temperature description of dipolar gases (in the same vein as the Zaremba-Nikuni-Griffin formalism of regular Bose-condensates~\cite{JacksonZaremba1,ZNGJournalOfLowTempPhys,ZNGBook}). This remains as {\it work-in-progress}. 

The method used in this paper (DSMC) is a remarkably versatile tool, potentially capable of simulating a multitude of out-of-equilibrium scenarios. Extending it into a regime where many-body quantum mechanical behaviour becomes prevalent (beyond the simple two-body scattering level which plays such a vital role in our current work) is a direction which we intend to take this research. Possible avenues for doing so include, incorporating the effects of Bose-stimulation and Pauli-blocking into the differential scattering cross sections, as prescribed by Nordheim~\cite{Nordheim}, see Eq.~\eqref{eq:Nordheim}. This requires modifications to the DSMC algorithm, which were originally introduced in the context of nuclear equations of state, particularly during heavy ion collisions~\cite{BoltzEqReview,Aichelin}. The basic ideas have seen application in ultra-cold atomic systems of fermions, see Refs.~\cite{LepersQuantumBoltzFermi,GoulkoFermiBoltz1}.   Another possibility, perhaps more relevant for bosonic systems, involves coupling the Boltzmann equation (the purely classical version may suffice) to an equation describing the superfluid component in the system. For example one could consider using the well-known Gross-Pitaevskii equation~\cite{ZNGBook}, or the more sophisticated $c$-field techniques~\cite{cfieldreview}.

\section{Acknowledgments}
AGS wishes to thank Andrew Wade and Blair Blakie for useful advice in developing the DSMC code. AGS and JLB both wish to thank Francesca Ferlaino and Kiyotaka Aikawa for sending us their experimental data, and providing useful feedback on our work. AGS and JLB also acknowledge interesting discussions with Benjamin Lev, Yijun Tang, Nathaniel Burdick, and Kristian Baumann regarding dysprosium gases.

\begin{appendix}

\section{Rejection sampling algorithm}\label{app:RejectionSampling}

The procedure of rejection sampling is not new~\cite{VonNeumannRejection}, but for completeness, we provide a brief description of the details specific to our situation. A more thorough description of the algorithm in general can be found in Ref.~\cite{MonteCarloBook}.

\subsection{Fermions}

To sample from $P_{\rm F}(\theta,\phi;\eta)$ defined in Eq.~\eqref{eq:ProbDistFun}, the strategy is to start from a simpler distribution (which is easy to sample), call it $g(\theta,\phi)=1/(2\pi^2)$, and (appropriately) reject those samples which were {\it unlikely} (recall that we only need to sample $\theta$ and $\phi$ since $\eta$ is given to us by the (already known) incoming relative momentum of the collision pair). The algorithm goes as follows:
\begin{enumerate}
 \item Sample $(\bar{\theta},\bar{\phi})$ from $g(\theta,\phi)$, and sample $u$ from $\mathcal{U}(0,1)$ (the uniform distribution over the unit interval).
 \item Check whether $u<P_{\rm F}(\bar{\theta},\bar{\phi};\eta)/[Mg(\bar{\theta},\bar{\phi})]$ where $M$ is an {\it upper-bound} such that $M>P_{\rm F}(\theta,\phi;\eta)/g(\theta,\phi)$ for all $\theta$ and $\phi$.
 \item If step 2 holds true, accept $(\bar{\theta},\bar{\phi})$ as a realisation of $P_{\rm F}$. If it does not hold true, reject $(\bar{\theta},\bar{\phi})$, and begin over at step 1.
\end{enumerate}

In order to find the upper bound $M(\eta)$ we transform to the collision-reference-frame, where
\begin{widetext}
\begin{equation}
 P_{\rm F}(\theta,\phi;\eta)=\frac{6\sin(\theta)\left[
\cos(\theta)\left(\cos^2(\eta)-\cos^2(\phi)\sin^2(\eta)\right)+\cos(\phi)\sin(\theta)\sin(2\eta)
\right]^2}
{\pi\left(3+18\cos^2(\eta)-13\cos^4(\eta)\right)}.\label{eq:PF}
\end{equation}
\end{widetext}
Using standard optimisation methods, we find the maximum value of $P_{\rm F}(\theta,\phi;\eta)$ occurs at $\phi_{\rm max}=0$, and
\begin{widetext}
\begin{equation}
 \theta_{\rm max}=\left\{\begin{array}{ll}
                          {\rm acos}\left(\frac{\sqrt{7+\cos(4\eta)-\sqrt{2
\sin^2(\eta)\left(17-\cos(4\eta)\right)
}}}{2\sqrt{3}} 
\right) & \eta<\pi/4 \quad\textrm{or}\quad \pi/2<\eta<3\pi/4\\
{\rm acos}\left(-\frac{\sqrt{7+\cos(4\eta)-\sqrt{2
\sin^2(\eta)\left(17-\cos(4\eta)\right)
}}}{2\sqrt{3}} \right) & \pi/4<\eta<\pi/2 \quad\textrm{or}\quad 3\pi/4<\eta.
                         \end{array}
\right.
\end{equation}
\end{widetext}
From this, we define $M=2\pi^2P_{\rm F}^{\rm (max)}(\eta)$ where,
\begin{equation}
 P_{\rm F}^{\rm (max)}(\eta)=\frac{6\cos^2(\theta_{\rm max}-2\eta)\sin(\theta_{\rm max})}
{\pi\left(3+18\cos^2\eta-13\cos^4\eta\right)}.
\end{equation}

\subsection{Bosons}
The procedure for bosons is essentially equivalent, except with;
\begin{widetext}
\begin{equation}
 P_{\rm B}(\theta,\phi;\eta)=\frac{2\sin(\theta)\left[-2+3\cos^2(\eta)+3\cos^2\left(\phi_{\rm max}\right)\sin^2(\eta)\right]^2}
{\pi\left(11-30\cos^2(\eta)+27\cos^4(\eta)\right)},\label{eq:PB}
\end{equation}
\end{widetext}
$\theta_{\rm max}=\pi/2$, 
\begin{equation}
 \phi_{\rm max}=\left\{\begin{array}{ll}
                          0 & \quad\eta<{\rm atan}\!\left(\sqrt{2}\right)  \quad\textrm{or}\quad \eta>\pi-{\rm atan}\!\left(\sqrt{2}\right)\\
\pi/2 & \quad {\rm atan}\!\left(\sqrt{2}\right)<\eta<\pi-{\rm atan}\!\left(\sqrt{2}\right).
                         \end{array}
\right.
\end{equation}
and
\begin{equation}
 P_{\rm B}^{\rm (max)}(\eta)=\frac{2\left[-2+3\cos^2\eta+3\cos^2\phi_{\rm max}\sin^2\eta\right]^2}
{\pi\left(11-30\cos^2\eta+27\cos^4\eta\right)}.
\end{equation}
Note that $P_{\rm B}$ in Eq.~\eqref{eq:PB} factorizes into a product of two functions involving only $\theta$ and only $\phi$. This was not the case for the fermionic cross section, see Eq.~\eqref{eq:PF}. This allows for the sampling algorithm to be more efficient in the case of bosons than it is for fermions, since $\theta$ can be sampled directly.
\end{appendix}

\end{document}